\documentclass[10pt]{article}
\usepackage{amssymb}
\usepackage{amsthm}
\usepackage{authblk}
\usepackage{setspace}
\usepackage[textwidth=6.5in,textheight=8.25in,centering]{geometry}
\usepackage[pdftex]{graphicx}

\onehalfspace
\begin{document}
\begin{center}
{\LARGE{\textbf{Size Dependent Sensitivity of Raman Line-Shape Parameters in Silicon Quantum Wire}}}

\vspace{0.5 cm}
\textit{Neeshu K.M. $^{a,b}$, Chanchal Rani $^{a,b}$, Ritika Kaushik$^{b}$, Manushree Tanwar$^{b}$, Ashisha Kumar$^c$ and Rajesh Kumar$^{b}$}\footnote{Corresponding author email: rajeshkumar@iiti.ac.in}

\vspace{0.5 cm}

$^a$ Authors have equal contribution

$^{b}$Discipline of Physics \& MEMS, Indian Institute of Technology Indore, Simrol-453552, India

$^c$ Discipline of Mathematics, Indian Institute of Technology Indore, Simrol-453552, India

\vspace{1 cm}
ABSTRACT

\end{center}
A comparison of experimentally observed Raman scattering data with Raman line-shapes, generated theoretically using phonon confinement model, has been carried out to understand the sensitivity of different Raman spectral parameters on quantum confinement effect. Size dependent variations of full width at half maximum (FWHM), Raman peak position and asymmetry ratio have been analyzed to establish the sensitivity of their corresponding physical counterparts (phonon life time and dispersion) in confined systems. The comparison has been done in three different confinement regimes namely, weakly, moderately and strongly. Proper reasoning has been assigned for such a variation after validation of the theoretical analysis with the experimental observations. A moderately confined system was created by preparing 6 nm sized Si NSs using metal induced etching. An asymmetrically broadened and red-shifted Raman line-shape was observed which established that all the parameters get affected in moderately confined system. Sensitivity of a given Raman spectral parameters has been shown to be used as a tool to understand the role of external perturbations in a material.      
\vspace{0.5cm}


\vspace{0.2cm}

\newpage
\section{Introduction}
Quantum confinement effect is one of the most significant effects in the current era of nanomaterials leading the research field in nanoscience and nanotechnology.  Raman spectroscopy, since its discovery in 1928, remains one of the most effective and easiest techniques when it comes to investigate quantum confinement effect [1, 2]. Raman spectroscopy provides a fast and convenient method to analyze the vibrational properties of solids [3-–8]. A Raman spectrum contains information about a solid through the response of solids’ phonons to the incident photons by means of scattering. In crystalline materials in the bulk form, the Raman scattering is limited to near zone centered phonons due to spectroscopic selection rule [9] and result in a symmetric Lorentzian line-shape spectrum peaked at a frequency corresponding to the zone center phonon. The width and symmetry of the spectrum, normally quantified through full width at half maximum or FWHM, depends (inversely) on the lifetime and dispersion respectively of the phonons involved in the scattering. Any deviation from a given value of the width, peak position (or shift in peak position) and/or asymmetry indicates perturbation of the phonon by any external condition. As an example, in nanostructures, asymmetrically broadened and red-shifted Raman spectra are observed because the general Raman selection rule gets relaxed and phonons other than the zone centered ones also contribute due to confinement of phonons in a crystallite of finite dimension [10-–13]. Above-mentioned fact about variation of different Raman parameters with respect to size is very simplistic way to put it as it is difficult to ascertain that all of the parameters (asymmetry, peak and broadening) will get affected equally or differently in a given confined system. It would be interesting to understand the behavior of each of these parameters and thus the physical quantity responsible for this, on the quantum confinement effect. 

Phonon confinement model (PCM) established by Richter et al. [14] and later modified by Campbell et al. [15] is the most successful attempt in understanding the non-symmetric Raman line-shape from polycrystalline solids including nanomaterials. According to the PCM, intensity of the first-order Raman scattering can be written as Eq. 1 below:

\begin{equation}
I(\omega) = \int _0 ^1 \frac{e^{-\frac{k^2L^2}{4a^2}}}{[\omega - \omega(k)]^2-(\gamma /2)^2}d^nk
\end{equation}

where, k is expressed in the units of 2/a, a being the lattice constant, 0.543 nm. The parameter L stands for the average size of the nanocrystals. γ being the linewidth of the Si optical phonon in bulk c-Si ($\sim$ 4 cm$^{−1}$) and `n’ shows the degree of confinement and may take values 2 or 3 for 2-dimensional or 3-dimensional confinement, respectively. Following Tubino et al. [16], the dispersion $\omega (k)$ of the optical phonon in a Si NSs can be taken as $\omega ^2(k)$ = 171400 + 100000 cos ($\pi$k/2). Equation 1 when compared with an experimentally observed Raman line-shape can give the size of the nanocrystal. However, Eq. 1 explains the observation of non-symmetric Raman line-shape due to quantum confinement effect, one can not ascertain the effect of quantum confinement effect on individual Raman line-shape parameters like peak shift, asymmetry ratio and peak width. It also cannot be ascertained that whether Bohr’s exciton radius can be treated as the universal confinement regime defining parameter in the context of Raman spectroscopy.

 Main aim of the present paper is to understand the sensitivity of different Raman spectral parameter towards the quantum confinement effect. This has been done by closely analyzing the response of Raman parameters, FWHM, peak position and asymmetry ratio to the crystallite size. Size dependent variations of FWHM, Raman peak position and asymmetry ratio have been analyzed to establish the sensitivity of their corresponding physical counterparts (phonon life time and dispersion) in confined systems. The analysis helps in establishing different confinement regimes where the different parameters start responding to the confinement effect. The phonon life time, manifested as FWHM, is observed to be most sensitive parameter which can be observed in the weakly confined systems also. An experimental validation of the abovementioned effect has also been done to establish the conclusions through taking the example of Si NSs.  
\section{Experimental Details}

A commercially available (Vin Karola$^{TM}$) Si wafers (n- type) with resistivity $\sim$0.1$\Omega$-cm has been used to fabricate the SiNSs using by metal induced etching (MIE) [17,18]. For MIE, first Ag nanoparticles (AgNPs)  were deposited on cleaned Si wafers by dipping them into a solution containing  4.8 M HF and 5 mM AgNO3 for 60 s at room temperature. The AgNPs deposited samples were then kept for etching in a solution containing 4.6 M HF and 0.5 M H2O2 for 45 minutes at room temperature. Etched wafers were then transferred to HNO3 to remove the AgNPs after the etching process. The samples were then dipped into HF solution to remove any oxide layer induced by nitric acid used in above step. A Supra55Zeiss scanning electron microscope (SEM) has been used to study the surface morphologies of these samples. Raman spectrum was recorded using a Horiba Jobin Yvon micro-Raman spectrometer with a 633nm excitation laser with minimum power to avoid any laser-induced heating.

\section{Results and Discussion}
The sample prepared after porosification of Si wafer using MIE [9] looks to have uniform porosity (Fig. 1a) with an average pore size of less than a micron. The porous looking morphology is actually a result of an assembly of well aligned wire like structures clubbed together as can be seen using the cross-sectional SEM images (inset, Fig. 1a). These pores are approximately a few tens of microns long as obtained after 45 min. of etching. These wires are known to be consisting of smaller nanostructures of sizes in the range of a few nanometers and are capable of exhibiting confinement [9] effects. Raman spectroscopy is one of the most suitable ways to investigate confinement effect especially that of phonons in nanostructures. A red-shifted and asymmetrically broadened Raman spectrum is treated as an indication of presence of confined phonons. The same has been observed from the present sample (discrete points, Fig. 1b) confirming the presence of phonon confinement effect in the Si NSs samples prepared using MIE. As mentioned above, the asymmetry, broadening and red shift are observed as a manifestation of confined phonons in a system and can be used to estimate the size of the nanoscrystals if these Raman parameters are analyzed carefully mainly within the framework of PCM [17,18]. The red shift, asymmetry and broadening are quantified through peak position ($\omega$), asymmetry ratio ($\alpha$) and full width at half maximum (FWHM or $\gamma$) respectively (defined in inset of Fig. 1b). All of these parameters can be matched with theoretically generated Raman line-shape function, given by Eq. 1 to estimate the nanocrystallite size. The fitting of the experimentally observed Raman scattering data (discrete blue points) with Eq. 1 (solid red line) has been shown in Fig. 2. The bet fit between the experimental and theoretical data yields a nanocrystallite size of 6 nm when a two dimensional confinement is considered. It is interesting here to observe that the deviation of $\omega$, $\alpha$ and $\gamma$ from those of the bulk counterparts are not equal in terms of their values apparently meaning that the effect of quantum confinement is not of same order on the three parameters and it is expected that one (or may be couple) of these will get affected more than and prior to the rest. To understand which of these gets affected first, the size dependent Raman line-shapes, generated by PCM, have been analyzed closely as discussed below.

\begin{figure}
\begin{center}
\includegraphics[width=16cm]{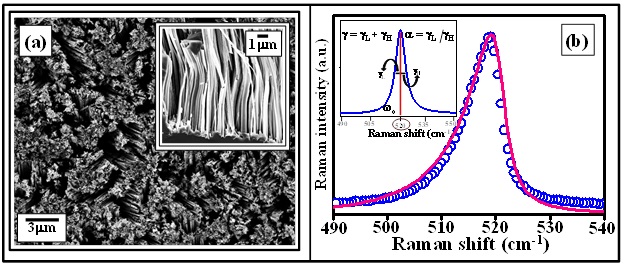}
\caption{(a) Surface morphology of Si NSs in top view and cross-sectional view (inset) and (b) Raman spectrum (discrete points) along with theoretical fitting (solid line) using phonon confinement model; inset shows various definitions of Raman parameters.}
\end{center}
\end{figure}

 Figure 2 shows the theoretical Raman line-shape function, predicted by PCM, as a function of nanocrystallite size (between 20 nm to 3 nm) confining the phonons in two dimensions (nanowires). It is very clear that the Raman line-shape, which is symmetric (with asymmetry ratio, $\alpha$, of unity) and centered at 521 cm$^{−1}$ with FWHM ($\gamma$) of ~ 4 cm$^{−1}$ for crystalline Si deviates from its bulk Raman line-shape on decreasing nanocrystallite size [19-23]. It is worth mentioning here that the Raman line-shape remains more or less like the bulk line-shape till the crystallite size as low as 20 nm which is four time the Bohr’s exciton radius value for Si hence showing little confinement effect. On further decreasing the size till 11 nm, the Raman line-shape starts showing some deviation from the bulk line-shape but this is visible only on very closer look (inset) as the effect of confinement on Raman line-shape parameters are rather very small. This is expected as this is the ``weak confinement” size regime. On further decreasing the sizes a rapid variation in line-shape can be appreciated when the sizes are comparable to 5 nm which also happens to be the Bohr’s exciton radius value. The Raman line-shape gets extremely asymmetrically broadened and red shifted with peak positioned at 515.5 cm$^{−1}$ carrying a FWHM of 35 cm$^{−1}$ and asymmetry ratio of 5 in the strong confinement regime for sizes ~ 3 nm. From Fig. 2, one can notice that different Raman line-shape parameters responds to the confinement effect differently in the different confinement regimes namely weakly, moderately and strongly confinement regimes. The same has been analyzed qualitatively below to extract any subtle physics taking place at these nanostructures. 
 
\begin{figure}
\begin{center}
\includegraphics[width=10cm]{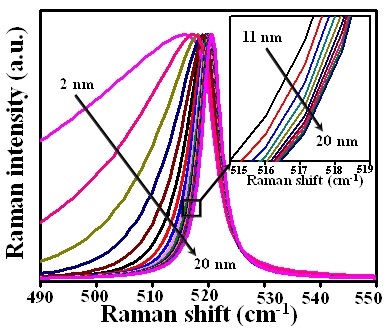}
\caption{Theoretical Raman line-shapes as a function of Si NSs size generated using phonon confinement model considering two-dimensionally confined system. Inset shows the zoomed portion of part of the main figure.}
\end{center}
\end{figure}

A size dependent variation of different Raman line-shape parameters (Fig. 3a) reveals that the three parameters ($\omega$, $\alpha$ and $\gamma$) vary in different ranges with the common size variation in the range of 20 nm to 3 nm. It is evident (Fig. 3a) that the variation by values (or the numbers) is maximum for FWHM which varies from 5 to 35 cm$^{−1}$ (seven times or 600\%) whereas the variation is minimum for peak position which varies from 521 to 515.5 cm$^{−1}$ (nearly $\sim$ 1\%). On the other hand, the asymmetry ratio vary moderately between the values 1 to 5 (five times or 400\%) while changing the size in the same above-mentioned range (20nm to 3 nm). Due to different variation regimes in the same size variation range it is not very easy to appreciate the effect quantitatively. For better understanding, derivatives of all the parameters ($\omega$, $\alpha$ and $\gamma$) have been taken and its variation as a function of size is shown in Fig. 3b. For better comparison, negative derivative has been used for peak position as it shows opposite behavior to that from asymmetry ratio and FWHM. Figure 3b clearly shows that FWHM varies most rapidly in comparison to the other two which vary nearly at the same rate as a function of size. It can be understood that though the confinement of same physical entity, phonons, its manifestation in $\omega$, $\alpha$ and $\gamma$is different. To understand the reason a closer analysis is required before any conclusion can be drawn. The FWHM is known to be the manifestation of phonon life time in a solid which means that it is the life time (of the confined phonons) which appears to be the most sensitive to the confinement effect. On the confinement scale, due to higher sensitivity, the FWHM starts responding to size variation as early as in the weakly confinement regime. On the other hand, the asymmetry and peak shift mainly depend on the nature of phonons’ dispersion relation. The effect of confinement on the peak position and asymmetry is not appreciable until the moderate to strong confinement regime is attained. This is the regime when the Raman selection rule is relaxed sufficiently so that the phonons sufficiently away from the zone center start taking part in Raman scattering. Another factor that affects $\omega$ and $\alpha$ is the slope of the phonon dispersion curve which, for Si is flat near the zone center. The regime after which the phonon dispersion curve’s flatness ceases, appreciable change in $\omega$ and $\alpha$ is visible of course compromised by the weak weighting factor term (Eq. 1). It is interesting to notice that simply looking variation of different Raman line-shape parameters ($\omega$, $\alpha$ and $\gamma$) in fact contains several subtle physics about the confined phonons at nanoscale including the phonon life time and phonon dispersion nature and degree of confinement.

\begin{figure}
\begin{center}
\includegraphics[width=16cm]{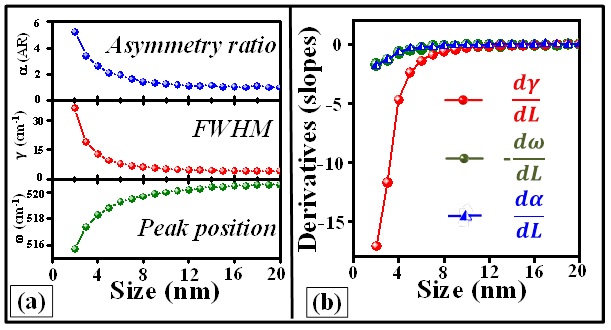}
\caption{(a) Variation in different Raman line-shape parameters, Raman peak position ($\omega$), FWHM ($\gamma$) and asymmetry ratio ($\alpha$) as a function of nanocrystallite size and corresponding rate of change of these parameters as a function of nanocrystallite size (b). Negative slope has been taken for peak position for better comparison (green curve, b).}
\end{center}
\end{figure}

\section {Conclusions}
An experimentally observed Raman scattering data and its comparison with theoretical phonon confinement model reveals that phonon life time is the most sensitive quantity to get affected by the confinement effect in Si NSs prepared by metal assisted etching. In terms of sensitivity, the relaxation of Raman scattering selection rule follows the phonon life time. In the one hand, the phonon life time starts getting affected even in the weakly confinement systems whereas the latter needs moderate to strong confinement effects to get manifested in the Raman line-shape. The above two observations is seen experimentally in following manner, a broadening (due to phonon life time) in the Raman line-shape starts appearing even in the regions where Raman peak shift and asymmetry ratio remains in the latent stages to be visible only in the strong confinement effect. An asymmetrically broadened and red-shifted Raman line-shape was observed from Si NSs of 6nm size (a moderate confinement regime) where variation in all the three parameters was observed which further validates the abovementioned conclusion. On the whole, sensitivity of a given Raman spectral parameters can be used to understand the role of external perturbations in a material.      
\section*{Acknowledgements} 
Authors acknowledge Sophisticated Instrumentation Centre (SIC), IIT Indore for SEM measurements. Financial Support from Department of Science and Technology (DST), Govt. of India is also acknowledged. Authors thank IIT Indore for providing fellowship. Facilities received from Department of Science and Technology (DST), Govt. of India, under FIST Scheme with grant number SR/FST/PSI-225/2016 is also acknowledged.

\newpage

\end{document}